\documentclass[conference]{IEEEtran}

\ifCLASSINFOpdf
\else
\usepackage[dvips]{graphicx}
\fi
\usepackage{url}

\usepackage{cite}
\usepackage{graphicx,amsmath,amssymb,tikz,psfrag}
\usepackage{color}
\usepackage{fancyhdr} 
\usepackage{tikz}
\usetikzlibrary{shapes,arrows,chains}
\usepackage{xcolor}
\usepackage{algorithm}
\usepackage{algpseudocode}

\begin{document}
	
	\title{Energy-Efficient UAV-Sensor Data Harvesting: Dynamic Adaptive Modulation  and Height Control}
	
\author{Dongsheng Chen, \textit{Graduate Student Member, IEEE},  \\

	\IEEEauthorblockA{Department of Electrical and Electronic Engineering, Southern University of Science and Technology, China }

}

	\maketitle
	
	\begin{abstract}
		Leveraging unmanned aerial vehicle (UAV) is convenient to collect data from ground sensor.
		However, in the presence of unknown urban environment, the data collection is subject to the blockage of urban buildings. In this paper, considering the urban environment during flight, we propose dynamic adaptive modulation and height control for UAV-sensor data harvesting in urban areas. In each time slot, the modulation format and flight height are selected based on current system states, with the aim of minimizing the expected transmission energy of sensor under data volume and flight height constraints. The dynamic adaptive modulation and height control problem is formulated as  constrained finite-horizon Markov decision processes (CMDP), which can be solved by backward induction algorithm. The advantage of proposed joint design over modulation selection only is illustrated via the computer simulations, where $48.23\%$ expected transmission energy can be saved for ground sensor.
	\end{abstract}
	

	\IEEEpeerreviewmaketitle

	\section{Introduction}
	
	Collecting data from ground sensors via unmanned aerial vehicle (UAV) has attracted a lot of interests in recent years. Due to the high maneuverability and flexibility of UAV, it is convenient to collect data from sensors via UAVs \cite{5}.  In urban environment, the abundant surrounding building incur unexpected blockage during air-ground communications. If the locations of buildings are known, we can plan the flight to avoid blockage and increase the communication efficiency. However, it is a cumbersome task to obtain all building locations. Hence, for convenience, the urban blockage is modeled as probability with area related parameters.


		On the other hand, MDP has been considered in resource allocation of wireless networks.
		For example, finite-horizon MDP was used to allocate power and transmission symbols in wireless caching network \cite{Lv2019,Lv2020}.
		In edge computing system, the offloading of computing tasks can be optimized via
		infinite-horizon MDP 
		\cite{Lyu2020,Huang2020}. In \cite{201}, a partially-observed MDP was applied in age-of-information-based scheduling. Nevertheless, for non-dynamic scheduling problem, the pure-binary integer programming was used in \cite{200}, where optimal performance is obtained.

	In this paper, we proposed an online modulation and flight height design for UAV-enabled data collection in urban areas with blockage. We consider a scenario that a fixed-wing UAV hovers over a sensor to collect data with a circle trajectory \cite{7}.  To minimize the transmission energy consumption of sensor under data volume and flight height constraints, the UAV needs to select the modulation scheme from a modulation set, and to decide to elevate, descend or maintain the height at each time slot. We use the Markov decision processes (MDP) approach to tackle this. By defining the MDP state, action, reward, and transition probability, the constrained optimization problem is transformed into finite-horizon MDP formulation, and then solved by backward induction algorithm.  Simulation results show that our proposed design performs better than the online modulation with fixed height design. This demonstrates the benefit of online flight height, i.e., the UAV can adjust the blockage probability and distance-dependent fading based on real-time realization. Simulation results also show that the performance enhancement by increasing the size of modulation set gets saturated if the set is sufficiently large.
	
	The organization of remaining letter is provided as follows: System model is depicted in Section-II. MDP approach is given in Section-III. Simulation is shown in Section-IV. We draw the conclusion in Section-V. 

	\section{System Model}
	
	\subsection{UAV-Sensor Data Harvesting}
	
\begin{figure}[!ht]
		\centering
		\includegraphics[width=2.75in]{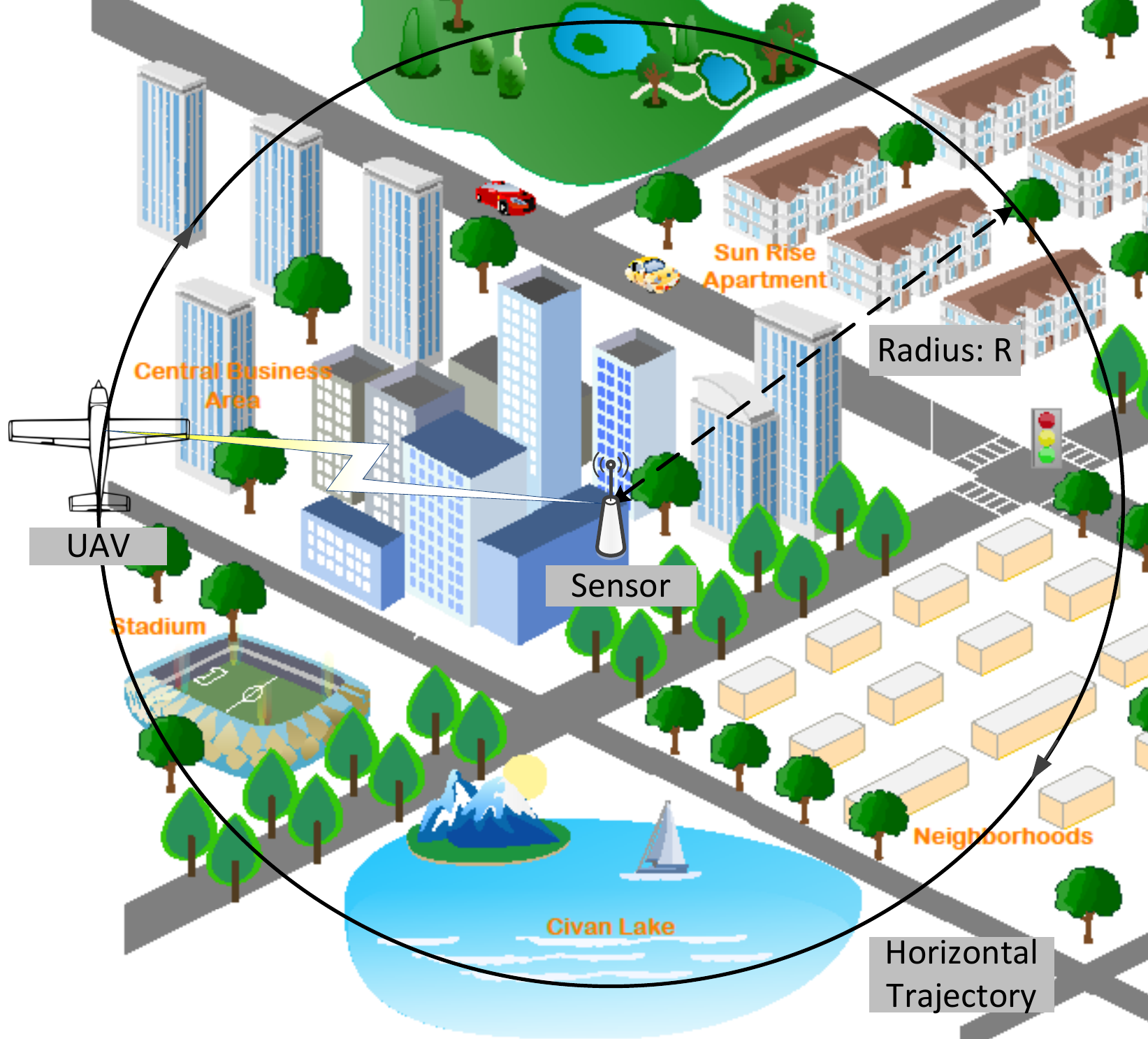}
		\caption{Illustration of UAV-enabled data collection in urban areas with the circular trajectory.}
		\label{Fig1}
	\end{figure}

	The considered  UAV-enabled data collection scenario is depicted in Fig. \ref{Fig1}, where a fixed-wing UAV collects a certain amount of data from a battery-limited sensor in an urban area. The UAV hovers over a sensor with a circle and a fixed speed.  Both the UAV and the sensor are equipped with single antenna.  The duration of flying is divided into $N$ time slots, where each time slot has $\tau$ duration. There is $D$ amount of data to  collect, such that bit error rate (BER) is required to not more than $\gamma$ and symbol rate is $r_s$\footnote{The research of fundamental limits of achievable rate can be found in \cite{100,101,102,103,104,105,106,107} and reference therein}.  at time slot $1$, the UAV just takes off and is at height $u$, and at time slot $N$ the UAV is about to land and at height $u$, where $u$ is the minimal distance of height adjustment. At each time slot, UAV needs to decide to elevate, maintain, or descend and select the modulation scheme from a modulation set. 
	
	Mathematically, the transmission and reception relationship at time slot $t$ is given by
	\begin{equation}
		y_t = \sqrt{g_t} x_t + n_t, \quad t \in \{1,\cdots,N\}
	\end{equation}
	where the transmit signal is denoted by $x_t$, the received signal is denoted by $y_t$, the additive white noise (AWGN) is denoted by  $n_t \sim {\cal{CN}}(0,\sigma^2)$, the path loss is denoted by  $g_t$.  The power of transmit signal $x_t$ is denoted by $p_t$.
	
	\subsection{Channel Model}
	The path loss is the attenuation of the transmitted signal to the received signal. According to \cite{2,9},  path loss can be modeled by the following widely-recognized elevation angle-dependent probabilistic line-of-sight (LoS) model:  
	\begin{equation}
		g_t = 
		\begin{cases}
			\beta_0 d_t^{-\alpha}, &  \text{LoS} \, (B_t = 0) \\
			{\cal{K}} \beta_0 d_t^{-\alpha}, & \text{NLoS} \, (B_t = 1) 
		\end{cases} \label{L}
	\end{equation}
	where $d_t^{-\alpha}$ denotes the distance-dependent fading at time slot $t$, $\beta_0$ is the path loss at the reference distance of 1m under LoS condition,  ${\cal{K}} < 1$ is the additional attenuation factor due to the NLoS propagation, and $B_t$ is a binary indicator for LoS or NLoS.  The probability of LoS is given by
	\begin{equation}
		\text{Pr}_{\text{LoS}}(\theta) = \frac{1}{1 + a\exp(-b(\theta - a))}
	\end{equation}
	where $a,b$ represent S-curve parameters, which can be directly to the environment variables, $\theta$ denotes the elevation angle. Since the distance between the UAV and the sensor at time slot $t$ can be calculated by 
	\begin{equation}
		d_t = \sqrt{H_t + R} \label{K}
	\end{equation} 
	where $H_t$ denotes the height at time slot $t$ and $R$ denotes the hovering radius. Since both the flight speed and sensor location are fixed, $R$ is a constant. Based on \eqref{K}, we can re-write the the elevation angle-dependent probabilistic LoS model in \eqref{L} to the following expression:
	\begin{equation}
		g_t = 
		\begin{cases}
			\beta_0 (H_t + R)^{-\alpha/2}, &  \text{LoS} \, (B_t = 0) \\
			{\cal{K}} \beta_0 (H_t + R)^{-\alpha/2}, & \text{NLoS} \, (B_t = 1) 
		\end{cases} 
	\end{equation}
	and the probability of LoS is given by
	\begin{equation}
		\text{Pr}_{\text{LoS}}(H_t)= \frac{1}{1 + a\exp(-b(\text{arctan}(H_t/R) - a))} \label{B}
	\end{equation}
	where we can see that, for a fixed $R$, elevating the flight height will increase the probability of LoS. 
	
	\subsection{Transmission Energy Consumption} 
	
	M-QAM adaptive modulation is adopted, because the flexible modulation selection has a better performance than the fixed modulation \cite{0,1}.  
	According to \cite{0,1}, the BER can be approximately calculated by 
	\begin{equation}
		\text{BER}  \approx 0.2\exp \left[ \frac{-1.6 g_t p_t}{\sigma^2(M_t  - 1)} \right]
	\end{equation}
	where $M_t \in \{1,2,\cdots\}$ denotes the selected modulation scheme. Note that we require that $M_t = 1$ represents transmission muting, and $M_t \ne 1$ represents $2^{M_t-1}$-QAM. Given the BER threshold $\gamma$, the corresponding energy consumption at time slot $t$ is given by
	\begin{equation}
		E_t= p_t\tau = \frac{\sigma^2(M_t - 1)\tau\ln (\gamma/0.2)}{-1.6 g_t}.
	\end{equation}

	\section{Problem Formulation and Solution}
	
	\subsection{MDP Problem Formulation}

	We use MDP approach to tackle this problem, since the MDP is a powerful mathematical approach to solve online problem \cite{8}.  We carefully define the MDP state, action, reward,  transition probability as follows:

	\textit{MDP State}:  A tuple of height $H_t$, remaining data to send $D_t$, and blockage indicator $B_t$.

	\textit{MDP Action}: A tuple of elevating-maintaining-descending variable $U_t \in \{u,0,-u\}$ and modulation variable $M_t \in {\cal{M}}$.

	\textit{MDP Reward}: The transmission energy consumption at time slot $t$, i.e., $E_t$.
	
	\textit{MDP Transition Probability}: A product of transition probability of $H_t$ to $H_{t+1}$, transition probability of $D_t$ to $D_{t+1}$, and transition probability of $B_t$ to $B_{t+1}$. In particular, we have
	\begin{equation}
		\text{Pr}(H_{t+1}|H_t) = \begin{cases}
			1, & H_{t+1} = H_t + U_t \\
			0,& \text{otherwise}
		\end{cases}
	\end{equation}
	and
	\begin{equation}
		\text{Pr}(D_{t+1}|D_t) = \begin{cases}
			1, & D_{t+1} = D_t - r_s \tau \log_2(M_t)\\
			0, & \text{otherwise}
		\end{cases}
	\end{equation}
	Due to the circle trajectory, we can assume that the blockage events are independent, i.e., $\text{Pr}(B_{t+1}|B_t) = \text{Pr}(B_{t+1})$. The probability of event $B_{t+1} = 0$ can be calculated by  \eqref{B}, and the  probability of event $B_{t+1} = 1$ is equal to $1 - \text{Pr}(B_{t+1} = 0)$.

	\textit{Constraints on MDP State}:  In addition, to deal with data volume and flight height constraints, we enforce two constraints on MDP state. 
	\begin{enumerate}
		\item \textit{Data Volume Constraint}:  $D_1 = D, D_{N+1} = 0$
		\item \textit{Flight height Constraint}:   $H_1 = u, H_{N} = u$
	\end{enumerate}

	Once the optimal action at each time slot is obtained, the offline lookup table can be formulated. We show the format of lookup table at time slot $t$ in Tab.I., where we enumerate all possibility of MDP states and present the optimal action for corresponding MDP states.  
	
	\subsection{Online Policy} 
	
	Via the design of modulation and flight height, we attempt to minimize the transmission energy consumption of sensor under data volume and flight height constraints. The aforementioned problem can be formulated as follows:
	\begin{eqnarray}
		\min_{\{U_t,M_t\}_{t=1}^N}&& \mathbb{E}\left\{\sum_{t=1}^N E_t\right\} \label{P1} \\
		\text{s.t.} && \text{BER} \le \gamma  \\
		&& \sum_{t=1}^N r_s \tau \log_2(M_t) = D \\
		&& H_1 = u, H_{N} = u  
	\end{eqnarray}
	As a solution of above problem, an offline lookup table can be designed accordingly, which contains all underlying situations and recommends the best policy.   
	\begin{table}
		\centering
		\caption{Lookup table at time slot $t$}
		\begin{tabular}{|c|c|c|c|}
			\hline 
			$\backslash$ & $(D_t = ,H_t =,B_t = )$ & $\cdots$ & $(D_t = , H_t = ,B_t = )$  \\ \hline 
			$(U_t^*,M_t^*)$ & & & \\ \hline 		 
		\end{tabular}
	\end{table}
	
	\begin{figure*}
		\begin{equation} \label{Bell} 
			V_i  = \min_{U_i,M_i} \left\{\underbrace{E_i}_{\text{current}} + \underbrace{\sum_{D_{i+1}}\sum_{H_{i+1}}\sum_{B_{i+1}} \text{Pr}(D_{i+1}|D_i)\text{Pr}(H_{i+1}|H_i)\text{Pr}(B_{i+1})V_{i+1}}_{\text{future}} \right\} 
		\end{equation}
		\hrule
	\end{figure*}

	\subsection{MDP Problem Solution} 
	To solve the above finite-horizon MDP,  we should define the value function for MDP state first, which is given as follows:
	\begin{equation}
		V_i  \triangleq  \sum_{t=i}^N E_t,
	\end{equation} 
	which is the cumulated rewards from time slot $i$ to time slot $N$ and function of MDP state. Therefore,  we needs to minimize the value function at initial state w.r.t. data volume and flight height constraints. This constrained finite-horizon MDP problem is written as follows:
	\begin{eqnarray}
		\min_{\{M_t,U_t\}_{t=1}^N} && V_1 \\
		\text{s.t.} && D_1 = D, D_{N+1} = 0 \\
		\, && H_1 = u, H_{N} = u
	\end{eqnarray}
	According to \cite{8}, the Bellman optimality equation is given in \eqref{Bell}. 
	The current value function can be represented by the summation of current reward and future reward. That is to say, if we know the value of $V_{i+1}$, we can derive the value of $V_i$. Because we can assign the value of $V_{N+1}$. Through Bellman equation, we must know the value of $V_1$ and corresponding the series of optimal actions for all time slots. The resultant algorithm is given in Algorithm 1, where the optimal action  is calculated in a backward manner and the constraints on MDP state are also considered.
	
	\begin{algorithm}
		\caption{Backward Induction Algorithm}
		\label{array-sum}
		\begin{algorithmic}[1]
			\State $i = N+1$: we set $V_{N+1} = 0$	for the case $D_{N+1} = 0$ and $H_{N+1} = 0$, and $V_{N+1} = +{\cal{1}}$ for other cases. 	 
			\State $i = N$:	 \begin{eqnarray}
				&& V_N  = \min_{U_N,M_N} \{E_N + \sum_{B_{N+1}} \text{Pr}(D_{N+1}=0|D_N) \nonumber \\
				&& \times \text{Pr}(H_{N+1}=0|H_N=u)\text{Pr}(B_{N+1})V_{N+1} \} \nonumber 
			\end{eqnarray} 
			\For {$i = N-1:-1:2$} 
			{\small\begin{eqnarray}
					&& V_i  = \min_{U_i,M_i} \{E_i +  \nonumber \\
					&& \sum_{D_{i+1}}\sum_{H_{i+1}}\sum_{B_{i+1}} \text{Pr}(D_{i+1}|D_i)\text{Pr}(H_{i+1}|H_i)\text{Pr}(B_{i+1})V_{i+1} \} \nonumber 
			\end{eqnarray}}
			\EndFor	
			\State  $i = 1$:
			{\small\begin{eqnarray}
					&& V_1  = \min_{U_1,M_1} \{E_1 +  \nonumber \\
					&& \sum_{D_{2}}\sum_{H_{2}}\sum_{B_{2}} \text{Pr}(D_{2}|D_1=D)\text{Pr}(H_{2}|H_1=u)\text{Pr}(B_{2})V_{2} \} \nonumber 
			\end{eqnarray}}	
		\end{algorithmic}
	\end{algorithm}

	\begin{table*} 
		\centering
		\caption{A Realization of Proposed Joint Modulation Selection and Flight Adjustment}
		\begin{tabular}{|c|c|c|c|c|c|c|c|c|c|c|}
			\hline 
			Time Slot & 1 & 2 & 3 & 4 & 5 & 6 & 7 & 8 & 9 & 10	\\ \hline 
			Blockage & \textit{No} & \textit{Yes} & \textit{No} & \textit{Yes} & \textit{No} & \textit{No} & \textit{Yes} & \textit{No} & \textit{Yes} & \textit{No} \\ \hline 
			Modulation & \textit{BPSK} & \textit{Muting} & \textit{BPSK} & \textit{Muting} & \textit{BPSK} & \textit{BPSK} & \textit{Muting} & \textit{BPSK} & \textit{Muting} & \textit{Muting}\\ \hline 
			height& \textit{30m} & \textit{60m} & \textit{90m} & \textit{120m} & \textit{120m} & \textit{120m} & \textit{120m} & \textit{90m} & \textit{60m} & \textit{30m} \\ \hline 		 
		\end{tabular}
	\end{table*}

	\section{Simulation}
	We examine the performance of proposed design via simulation. The setting used throughout the simulation are given as follows: The total fight time is $500$s, which is divided into $10$ time slots. The radius of circle trajectory is $50$m. The power density of AWGN is $-120$dBm/Hz.  The BER threshold is $10^{-5}$. For path loss model, we set $\alpha = 3$, $\beta_0 = 1$, ${\cal{K}} = 10^{-3}$, and $a=b=1$. Without loss of generality, we assume that there is no blockage in the initial state, i.e., LoS.

	Tab. II shows a realization of proposed online modulation and Flight height design, where the modulation set is $\{\text{Muting},\text{BPSK}\}$ and the minimal distance of height adjustment is $30$m. $30$Mbits data needs to be transmitted by sensor. Tab. II shows that 
	the UAV elevates the height in the early stage to avoid blockage, maintains the height in the medium stage to balance the blockage probability and distance-dependent fading, and descends the height in the late stage for landing.  Tab. II also shows that the UAV begin to transmit with BPSK modulation when there is no blockage in this time slot, and mutes when there is a blockage in this time slot.
	
\begin{figure}[!t]
		\centering
		\includegraphics[width=3.2in]{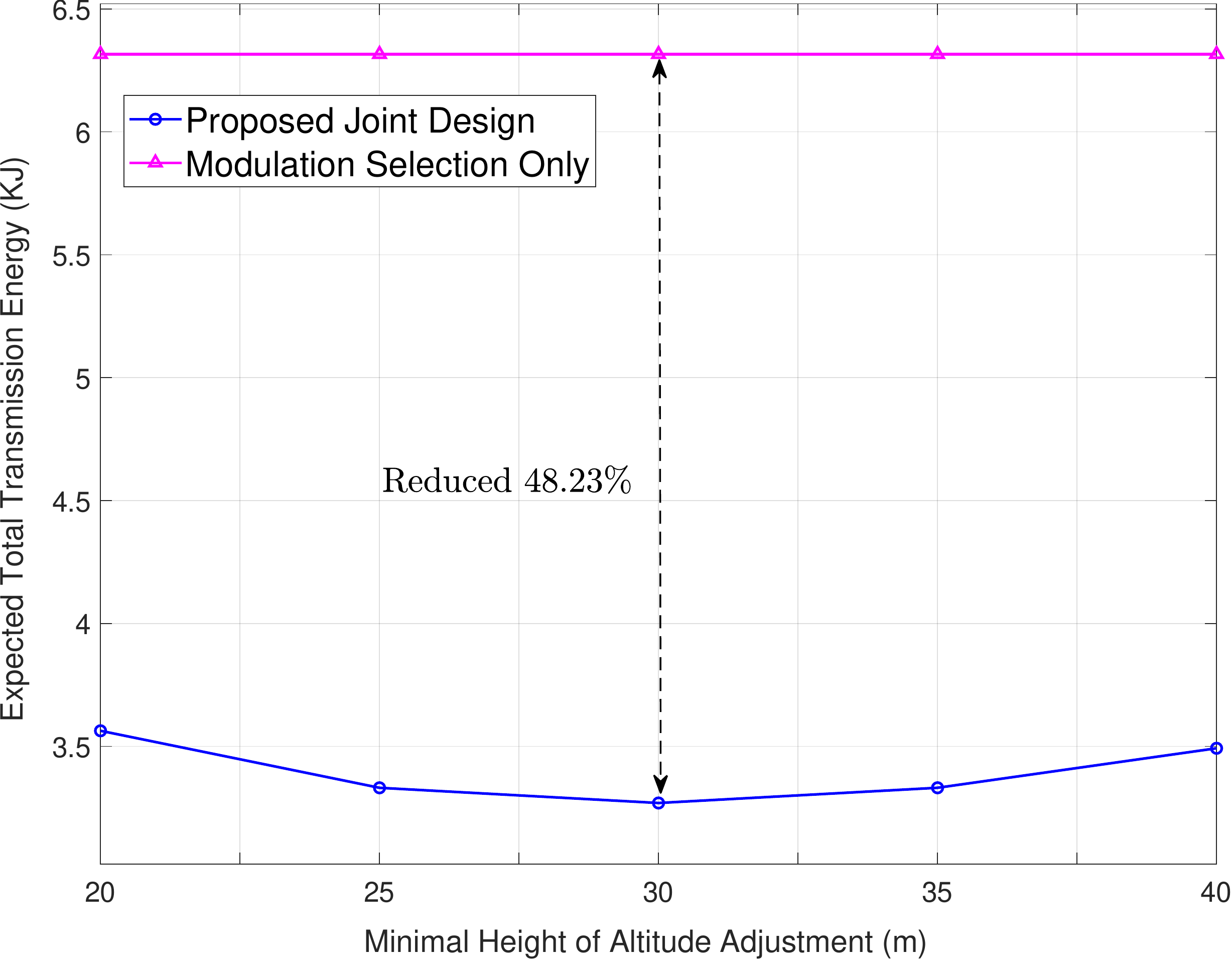}
		\caption{Proposed joint design v.s. modulation selection only.}
		\label{F1}
	\end{figure}

	In Fig. \ref{F1}, we compare the proposed online modulation and flight height design with online modulation and fixed height design. We set that $30$Mbits data needs to be transmitted by sensor and the modulation set is $\{\text{Muting},\text{BPSK}\}$. For online modulation and fixed height design, we enumerate all heights from $1$m to $400$m and find the optimal one, which is $23$m. We refer this as optimal fixed height design. Fig. \ref{F1} shows that the proposed online modulation and flight height design has a prominent advantage over the optimal fixed height design. The gain is varying w.r.t. minimal distance of height adjustment $u$. Fig. \ref{F1} shows that the gain is maximal when $u=30$m, where the maximal transmission energy consumption reduction is shown to be $48.23\%$. The advantage of proposed design over fixed height design comes from a fact that UAV can change the blockage probability and distance-dependent fading through adjusting the flight height. This is, in real time, UAV can elevate the height to reduce the blockage probability when there are many blockage events, or descend the height to reduce the distance-dependent fading when there are few blockage events.

\begin{figure}[!t]
		\centering
		\includegraphics[width=3.3in]{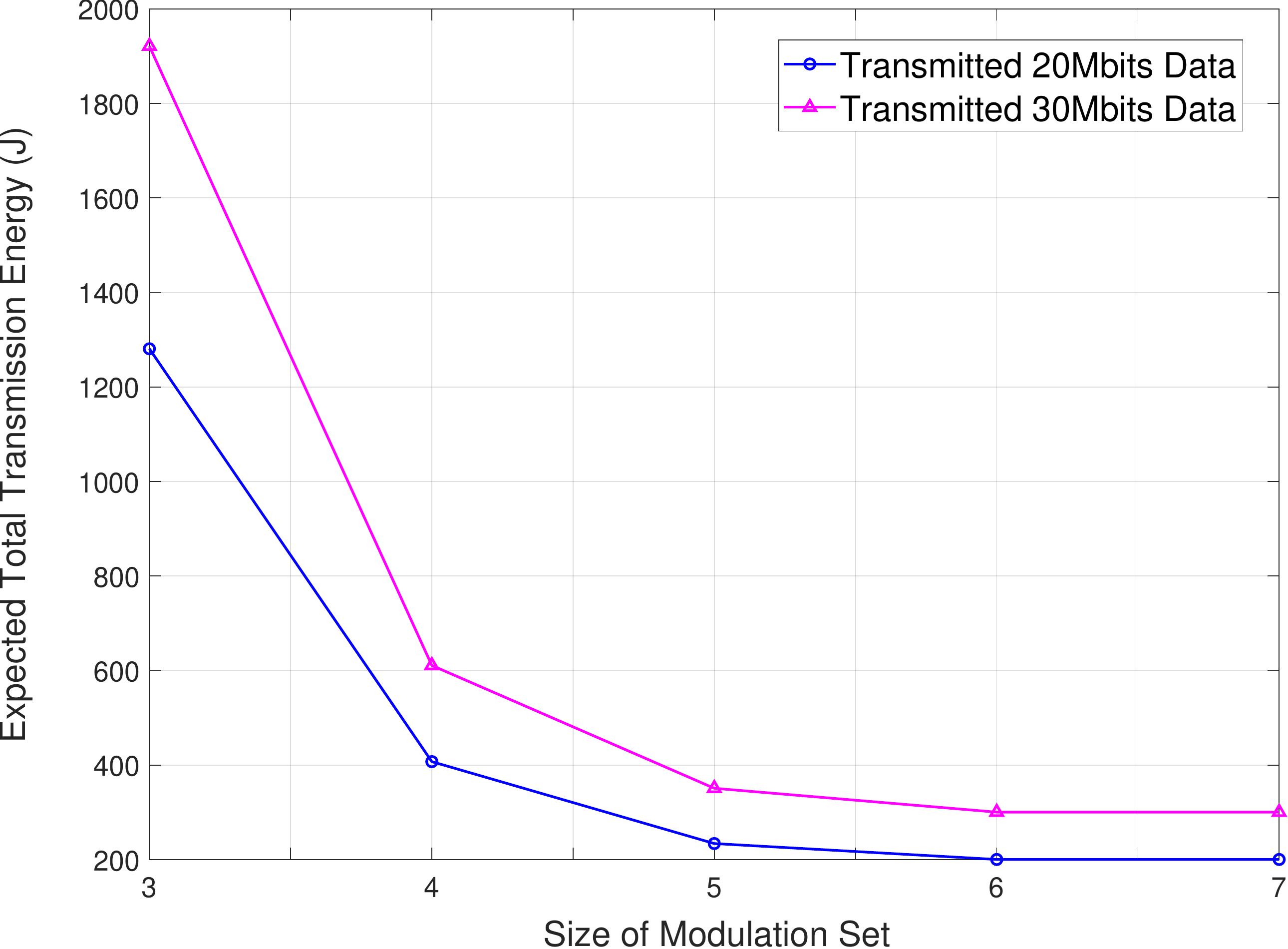}
		\caption{The impact of size of modulation set.}
		\label{F2}
	\end{figure}

	In Fig. \ref{F2}, we compare the performance of different size of modulation set to examine the impact of size of modulation set. The high-order modulation performs better in mild path loss and the low-order modulation is more preferable in severe path loss \cite{0,1}, hence it is necessary to adjust the modulation scheme based on real-time situations. Fig. \ref{F2} shows that enlarging the size of modulation set will not  always reduce the total transmission energy consumption. In this setting, the performance enhancement gets saturated if the size of modulation set is more than $6$, i.e., modulation set is $\{\text{Muting},\text{BPSK},\text{4-QAM},\text{8-QAM},\text{16-QAM},\text{32-QAM}\}$.
	This is because, higher-order modulations need an unacceptable transmission energy, thus will not be adopted even if they can provide a higher data rate. 
	
	\section{Conclusion}
	
	We proposed an online modulation and flight height design for UAV-enabled data collection in urban areas with blockage. Via MDP approach, the proposed design can acquire minimal sensor transmission energy consumption, through adjusting the modulation scheme, deciding to elevate, descend, or maintain the height, in an online manner. Simulation results demonstrate the benefit of online flight height through comparison with fixed height design. Simulation results also show that the performance enhancement by increasing the size of modulation set gets saturated if the set is sufficiently large.

	\bibliographystyle{IEEEtran}
	\bibliography{UAVSensor}

\begin{thebibliography}{10}
\providecommand{\url}[1]{#1}
\csname url@samestyle\endcsname
\providecommand{\newblock}{\relax}
\providecommand{\bibinfo}[2]{#2}
\providecommand{\BIBentrySTDinterwordspacing}{\spaceskip=0pt\relax}
\providecommand{\BIBentryALTinterwordstretchfactor}{4}
\providecommand{\BIBentryALTinterwordspacing}{\spaceskip=\fontdimen2\font plus
\BIBentryALTinterwordstretchfactor\fontdimen3\font minus
  \fontdimen4\font\relax}
\providecommand{\BIBforeignlanguage}[2]{{%
\expandafter\ifx\csname l@#1\endcsname\relax
\typeout{** WARNING: IEEEtran.bst: No hyphenation pattern has been}%
\typeout{** loaded for the language `#1'. Using the pattern for}%
\typeout{** the default language instead.}%
\else
\language=\csname l@#1\endcsname
\fi
#2}}
\providecommand{\BIBdecl}{\relax}
\BIBdecl

\bibitem{5}
Y.~{Zeng}, R.~{Zhang}, and T.~J. {Lim}, ``Wireless communications with unmanned
  aerial vehicles: Opportunities and challenges,'' \emph{IEEE Communications
  Magazine}, vol.~54, no.~5, pp. 36--42, May 2016.

\bibitem{Lv2019}
B.~Lv, L.~Huang, and R.~Wang, ``Joint downlink scheduling for file placement
  and delivery in cache-assisted wireless networks with finite file lifetime,''
  \emph{IEEE Transactions on Communications}, vol.~67, no.~6, pp. 4177--4192,
  June 2019.

\bibitem{Lv2020}
B.~Lv, R.~Wang, Y.~Cui, Y.~Gong, and H.~Tan, ``Joint optimization of file
  placement and delivery in cache-assisted wireless networks with limited
  lifetime and cache space,'' \emph{IEEE Transactions on Communications},
  vol.~68, no.~4, pp. 2339--2354, April 2020.

\bibitem{Lyu2020}
B.~Lyu, Y.~Hong, H.~Tan, Z.~Han, and R.~Wang, ``Cooperative jobs dispatching in
  edge computing network with unpredictable uploading delay,'' \emph{Journal of
  Communications and Information Networks}, vol.~5, no.~1, pp. 75--85, March
  2020.

\bibitem{Huang2020}
S.~Huang, B.~Lv, R.~Wang, and K.~Huang, ``Scheduling for mobile edge computing
  with random user arrivals—an approximate mdp and reinforcement learning
  approach,'' \emph{IEEE Transactions on Vehicular Technology}, vol.~69, no.~7,
  pp. 7735--7750, July 2020.

\bibitem{201}
A.~Gong, T.~Zhang, H.~Chen, and Y.~Zhang, ``Age-of-information-based scheduling
  in multiuser uplinks with stochastic arrivals: A {POMDP} approach,'' in
  \emph{GLOBECOM 2020 - 2020 IEEE Global Communications Conference}, 2020, pp.
  1--6.

\bibitem{200}
T.~Zhang, X.~Tao, and Q.~Cui, ``Joint multi-cell resource allocation using pure
  binary-integer programming for {LTE} uplink,'' in \emph{2014 IEEE 79th
  Vehicular Technology Conference (VTC Spring)}, 2014, pp. 1--5.

\bibitem{7}
Y.~{Zeng} and R.~{Zhang}, ``Energy-efficient {UAV} communication with
  trajectory optimization,'' \emph{IEEE Transactions on Wireless
  Communications}, vol.~16, no.~6, pp. 3747--3760, June 2017.

\bibitem{100}
T.~Zhang, S.~Wang, T.~Wang, and R.~Wang, ``The {DoF} region of order-({K-1})
  messages for the {K}-user {MIMO} broadcast channel with delayed {CSIT},'' in
  \emph{2021 IEEE/CIC International Conference on Communications in China
  (ICCC)}, 2021, pp. 688--693.

\bibitem{101}
T.~Zhang and R.~Wang, ``Secure degrees-of-freedom of the {MIMO} {X} channel
  with delayed {CSIT},'' \emph{IEEE Wireless Communications Letters}, vol.~10,
  no.~6, pp. 1319--1323, 2021.

\bibitem{102}
T.~Zhang, Y.~Xu, S.~Wang, M.~Wen, and R.~Wang, ``On secure degrees of freedom
  of the {MIMO} interference channel with local output feedback,'' \emph{IEEE
  Internet of Things Journal}, vol.~8, no.~20, pp. 15\,334--15\,348, 2021.

\bibitem{103}
T.~Zhang and R.~Wang, ``Achievable {DoF} regions of three-user {MIMO} broadcast
  channel with delayed {CSIT},'' \emph{IEEE Transactions on Communications},
  vol.~69, no.~4, pp. 2240--2253, 2021.

\bibitem{104}
T.~Zhang and P.~C. Ching, ``Secure {MIMO} interference channel with
  confidential messages and delayed {CSIT},'' in \emph{ICASSP 2019 - 2019 IEEE
  International Conference on Acoustics, Speech and Signal Processing
  (ICASSP)}, 2019, pp. 2437--2441.

\bibitem{105}
T.~Zhang, X.~Wu, Y.~Xu, Y.~Ge, and P.~C. Ching, ``Three-user {MIMO} broadcast
  channel with delayed {CSIT}: A higher achievable {DoF},'' in \emph{2018 IEEE
  International Conference on Acoustics, Speech and Signal Processing
  (ICASSP)}, 2018, pp. 3709--3713.

\bibitem{106}
T.~Zhang and P.~C. Ching, ``Interference alignment on {MIMO} {X} channel with
  synergistic {CSIT},'' in \emph{2017 IEEE International Conference on
  Acoustics, Speech and Signal Processing (ICASSP)}, 2017, pp. 3754--3758.

\bibitem{107}
\BIBentryALTinterwordspacing
T.~Zhang, G.~Chen, S.~Wang, and R.~Wang, ``Full-duplex relay with delayed {CSI}
  elevates the {SDoF} of the {MIMO} {X} channel,'' \emph{Entropy}, vol.~23,
  no.~11, 2021. [Online]. Available:
  \url{https://www.mdpi.com/1099-4300/23/11/1484}
\BIBentrySTDinterwordspacing

\bibitem{2}
A.~{Al-Hourani}, S.~{Kandeepan}, and S.~{Lardner}, ``Optimal {LAP} altitude for
  maximum coverage,'' \emph{IEEE Wireless Communications Letters}, vol.~3,
  no.~6, pp. 569--572, Dec 2014.

\bibitem{9}
M.~{Mozaffari}, W.~{Saad}, M.~{Bennis}, and M.~{Debbah}, ``Unmanned aerial
  vehicle with underlaid device-to-device communications: {Performance} and
  tradeoffs,'' \emph{IEEE Transactions on Wireless Communications}, vol.~15,
  no.~6, pp. 3949--3963, June 2016.

\bibitem{0}
A.~J. {Goldsmith} and {Soon-Ghee Chua}, ``Variable-rate variable-power {MQAM}
  for fading channels,'' \emph{IEEE Transactions on Communications}, vol.~45,
  no.~10, pp. 1218--1230, Oct 1997.

\bibitem{1}
{Seong Taek Chung} and A.~J. {Goldsmith}, ``Degrees of freedom in adaptive
  modulation: a unified view,'' \emph{IEEE Transactions on Communications},
  vol.~49, no.~9, pp. 1561--1571, Sep. 2001.

\bibitem{8}
M.~L. Puterman, \emph{Markov Decision Processes: {Discrete} Stochastic Dynamic
  Programming}.\hskip 1em plus 0.5em minus 0.4em\relax John Wiley \& Sons,
  2014.

\end{thebibliography}
	
\end{document}